\documentclass[conference]{IEEEtran}
\IEEEoverridecommandlockouts
\usepackage{verbatim}
\usepackage{textcomp}
\usepackage{subcaption}
\usepackage{amsmath,amssymb,amsfonts}
\usepackage{units}
\usepackage[font=small,skip=5pt,size=footnotesize,position=bottom]{caption}
\usepackage{xcolor}
\usepackage{cite}
\ifCLASSINFOpdf
\else
\fi
\usepackage{url}

\usepackage{graphicx}
 \usepackage[ruled,vlined]{algorithm2e}

\begin{document}
%
\title{Measurement-based Condition Monitoring of Railway Signaling Cables}


\author{\IEEEauthorblockN{Rathinamala Vijay\IEEEauthorrefmark{1},
Gautham Prasad\IEEEauthorrefmark{2}, 
Yinjia Huo\IEEEauthorrefmark{2},
SM Sachin\IEEEauthorrefmark{1}, and
TV Prabhakar\IEEEauthorrefmark{1}}
\IEEEauthorblockA{\IEEEauthorrefmark{1}Indian Institute of Science, Bangalore, India}
\IEEEauthorblockA{\IEEEauthorrefmark{2}The University of British Columbia, Vancouver, BC, Canada}}
%




\maketitle

\IEEEdisplaynontitleabstractindextext

\begin{abstract}
We propose a composite diagnostics solution for railway infrastructure monitoring. In particular, we address the issue of soft-fault detection in underground railway cables. We first demonstrate the feasibility of an orthogonal multitone time domain reflectometry based fault detection and location method for railway cabling infrastructure by implementing it using software defined radios. Our practical implementation, comprehensive measurement campaign, and our measurement results guide the design of our overall composite solution. With several diagnostics solutions available in the literature, our conglomerated method presents a technique to consolidate results from multiple diagnostics methods to provide an accurate assessment of underground cable health. We present a Bayesian framework based cable health index computation technique that indicates the extent of degradation that a cable is subject to at any stage during its lifespan. We present the performance results of our proposed solution using real-world measurements to demonstrate its effectiveness.

\end{abstract}


%
\IEEEpeerreviewmaketitle

\section{Introduction}
%
%
%
%
\IEEEPARstart{C}{able} diagnostics plays a crucial role in realizing the concept of the smart grid. Toward this end, several solutions have been proposed in the past for monitoring the health of power cables and detecting, locating, and assessing possible degradations and faults along the cable~\cite[Ch. 6]{gill2016electrical},~\cite{huo2018grid, prasad2019fault, jiang2011hybrid}. While most works in the literature focus, rightly so, on diagnosing cable health in the electricity transmission and distribution infrastructure, methods developed for these domains may not be universally applicable. The network setting considered in this paper is one such example, where we focus on cable diagnostics for railway signaling cables~\cite{underground}. For the purposes of this study, we focus on an Indian railway network setting  for example, considering network topology, operating loads, and cable types. However, the solutions we propose in this paper are generally universal in nature and can be extended across different operating conditions. Railway signaling cables are typically buried underground and carry power and signalling data for circuits such as: (a) point machine and the associated relay circuit that carry signalling data, (b) train-on-track circuit, (c) signalling speed limit indication circuit, which comprises of indications such as a yellow light (low speed), a double yellow light (much lower speed), and STOP and GO signal using red and green lights, respectively. These vital signals are used by the railway personnel and the loco pilot for the safe operation of trains. The power and control signal carrying cables are thus one of the critical signaling assets. 

The standard railway cables are 10 Twisted Pairs (TWP) and 6 quad Polyethylene Insulated Jelly Filled (PIJF).The PIJF insulation makes these cables suitable for sub-soil deployment. Cables buried underground are typically subject to insulation faults that are hard to detect by visual inspection~\cite{densley2001ageing}. Such faults can occur due to mechanical stress, accidental damage, and/or electrical stress. An electrical stress is usually caused due to short circuit in the loads (signalling loads in our case), and might lead to the creation of cable hotspots. These hotspots are potential locations for an insulation fault, which are as a result of ineffective heat dissipation, leading to rise in temperature. Based on the length of the fault, one may classify insulation fault as a small fault (e.g., 5mm to 3 cm insulation damage) or a large fault (e.g., 3 cm or longer insulation damage).

Cable faults are dangerous in most contexts of the smart grid~\cite{reichl2013value, castillo2014risk}. However, they are potentially more critical in the context of railway signaling cables. For example, the EU-27 has recorded about $802$ persons killed and $612$ persons injured during the year 2019 in Europe~\cite{eufatal}. The majority of the causes for these accidents are due to signaling faults. In another incident, three Indian railway staff deaths during a maintenance regime was reported in the year 2020~\cite{rail_staff}.     

The state-of-the-art in railway cable maintenance and diagnostics body is considerably lagging to other diagnostics counterparts. Currently, the cable maintenance has a periodic schedule where the insulation resistance of the cable is measured using a \textit{megger} test \cite{megger}. This measurement schedule is carried out once a year. Often, this schedule is complemented by railway personnel carrying out periodic visual inspection of the cable route to identify any recent digging activity or change in soil compaction~\cite{maintenance}.

IoT technologies can potentially fill the gap with effective real time  monitoring methods that can continuously look for anomalies. The cabling infrastructure connected to the point machine and the signaling and train on track detection circuit is expected to be monitored continuously for faults.  Such predictive maintenance not only enhances the passenger and train safety, but also reduces  downtime. The need for IoT platforms for railway asset management and the advantages from an economy and safety perspective is also brought out in the literature, e.g.,~\cite{gbadamosi2021iot}
\begin{figure}
     \centering
     \includegraphics[width=0.9\columnwidth]{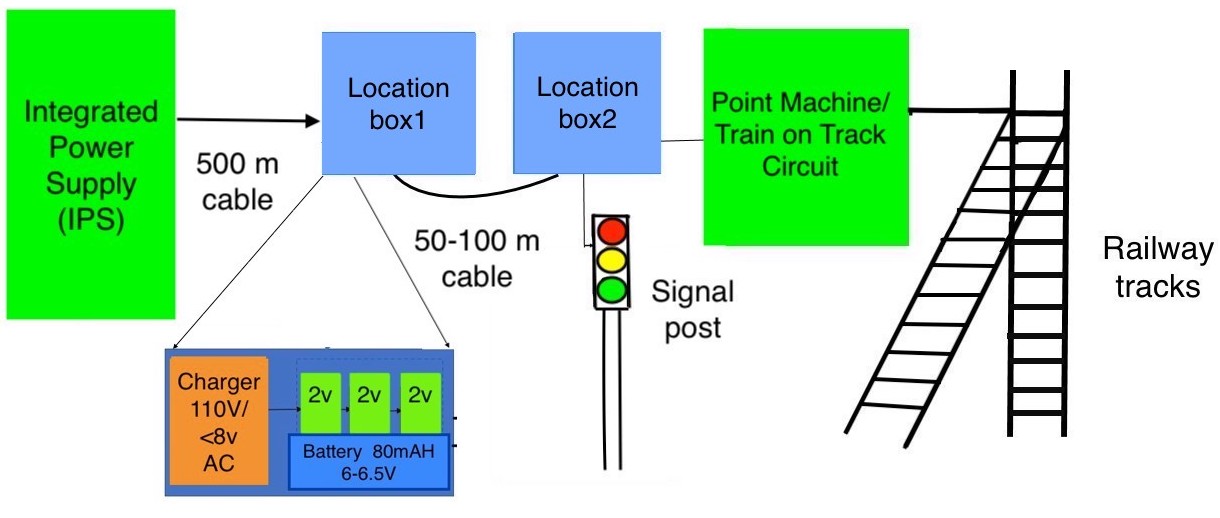}
     \caption{Real-time monitoring of a 500 m cable section between IPS and Location boxes is important. The location box provides uninterrupted power to the train on track detection circuit, the Point Machine circuit and Signal post circuits. A 50-100 m cable is used between any two Location boxes. Monitoring of these cable sections for insulation faults is crucial for a sustainable operation of train traffic flows.}
     \label{Chm_track}
     \vspace{-15 pt}
 \end{figure}

The problem we address in this paper is \textit{to reliably detect and localize a cable fault and thus compute a  health index for a cable}. Our  early fault detection system can be deployed between the Integrated Power Supply  and the location boxes. We develop an IoT solution for detecting, locating  and classifying cable faults. The network topology considered here is a single $500$~m long cable with no branches,  with a set of fixed and known loads applied at the location box as shown in Fig \ref{Chm_track}. Our laboratory scale setup and experiments mimic the actual railway cable section under test.

Our contributions can be listed as follows:
\begin{itemize}
    \item We implement a powerline testbed of a portion of a railway cabling infrastructure for continuous monitoring, detection, and localisation of cable faults. The system encompasses measurement based evaluation, incorporating (a) Signal to Noise Ratio (SNR) based tests (b) S-parameter based tests and (c) OMTDR based reflectometry tests. 
    \item We develop a Bayesian framework based multi-pronged consolidation solution that combines various sensing modalities to reliably identify small and large cable faults.
    \item We develop and evaluate a health index metric for  different cable types that indicate the severity of a possible fault. The health index provides an indication into the remaining useful life in the cable.
\end{itemize} 
\section{Related work}
 One of the earliest reflectometry  techniques for  cable fault detection and localization is Time Domain Reflectometry (TDR). Spread Spectrum Time Domain Reflectometry (SSTDR)~\cite{smith2005analysis,alam2013pv}, Orthogonal multi-tone Reflectometry (OMTDR)~\cite{hassen2019shielding}, and few other time domain reflectometry are used for fault detection and localization techniques~\cite{furse2020fault}. In general, Reflectometry methods are based on cross-correlation between the transmit test signal and received reflected signal. Although all these works demonstrate fault detection and localization, OMTDR has the ability to choose the test transmit signal frequencies such that the system under operation is not affected.   
 
 The authors of paper \cite{NASA2011} state that the S parameter measurement based cable fault detection is significantly more reliable than correlation based methods. The authors use Bayesian framework to estimate the fault characteristics. While we have used a similar approach to estimate fault characteristics, we additionally provide an overall health index metric for the cable. This metric gives an accurate knowledge of the cable health.
 
 The authors of  (\cite{prasad2019fault,huo2018cable}) use channel frequency response and SNR estimates from PLC modem and augumented machine learning to identify thermal degradation and short-circuit faults respectively in powerline cables through simulation studies. Although, these works prove by fault detection over simulations, we have built one of our testbeds with  commercial off-the-shelf (COTS) PLC modem and evaluated the soft fault identification accuracy with random load conditions. 
 
 The authors in \cite{boler2019novel} have investigated the fault in DC overhead railway traction power lines. In contrast to this work, we investigate underground cables which have a separate set of  challenges such as being prone to contamination of foreign materials, corrosion due to water clogs. \cite{grzechca2021effect} investigates the outer insulation degradation in shielded twisted pair railway cables and its impact on communication parameters. Our testbeds demonstrate a non-invasive online fault detection and location. Even though some of the techniques mentioned in the literature are capable of online measurements, placement of sensors have to be planned ahead and installed within a connector at a fixed position. In contrast, our non-invasive prototype can be used to test at random points on  the cable when the system is under normal operation. This flexibility reduces the downtime for maintenance schedule. 
\begin{figure}
     \centering
     \includegraphics[width=0.85\columnwidth]{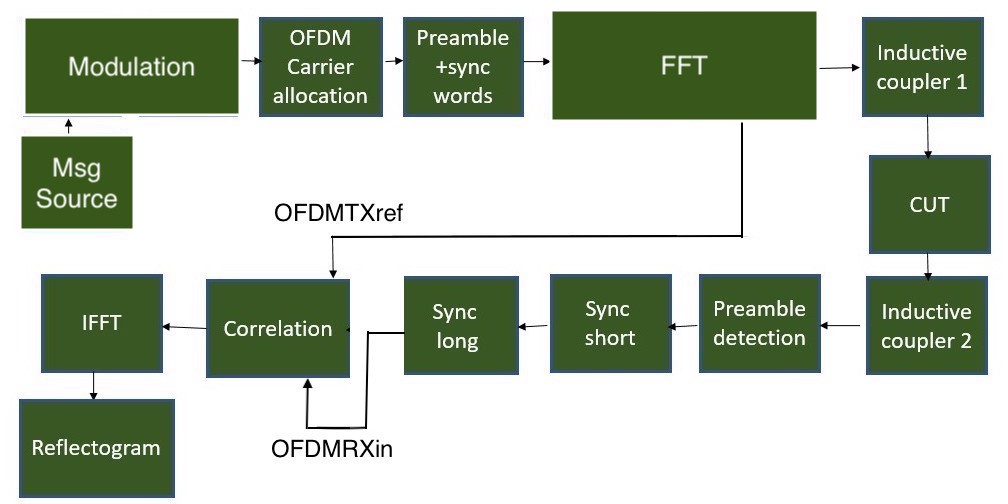}
     \caption{OMTDR basic blocks showing test signal injection in cable under test and the Reflectogram generation.}
     \label{fig_corr}
     \vspace{-15 pt}
 \end{figure}
\section{Fault detection methods}
We describe here three different fault detection methods namely OMTDR based Reflectometry test, S parameter test and SNR test. We then combine these three methods to improve the reliability of fault detection, thus leading to a useful  predictive maintenance metric. 

\begin{figure}
     \centering
      \includegraphics[width=0.9\columnwidth]{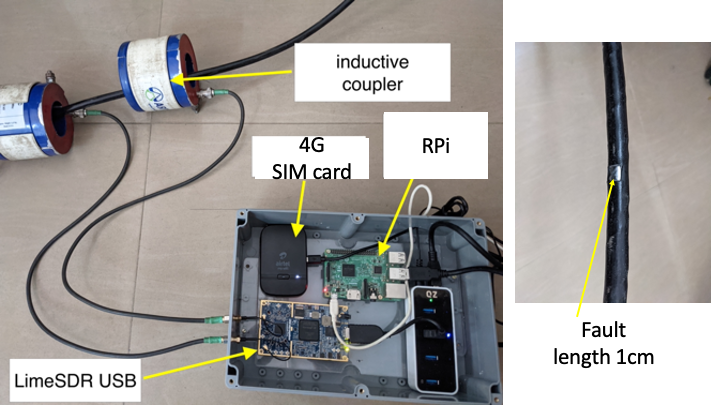}
     \caption{The OMTDR sensor constructed using the LimeSDR, a SDR module. The sensor injects an RF signal that couples to the cable using inductive couplers. The figure also shows the picture of 10TWP PIJF cable with insulation fault.  The sensor, RPi and 4G router form a typical system for cloud connectivity; irrespective of the sensing modalities.}
     \label{fig:exp_testbed}
     \vspace{-15 pt}
 \end{figure}

Table \ref{tab:testbed_comp} lists the testbed components and associated specifications.  The Raspberry PI computer, 4G router and LimeSDR (USB) together consume about 39 watts. The 4G router provides the required cloud connectivity.  The inductive couplers provide non-invasive coupling of test signals required for cable fault identification and localisation. This enables us to monitor the cables without disturbing the system under operation. The sample cables used for measurements is also shown.  The VNA and PLC modem are primarily used as sensors for S parameter and SNR measurements respectively. We perform exhaustive measurements under varying load conditions. Our loads  try to emulate the power requirements for signal posts, train on track and the point machine.  We basically use 3 lamps connected in parallel, where the load power can be varied from 200W to a maximum of 600W. Fig \ref{fig:exp_testbed} shows the setup for connecting a Cable Under Test (CUT) to 110 V 50 Hz AC supply. The train on track detection and signal circuits uses the Power Supply (PS) through the 6 quad and 10 TWP PIJF cables. A 10 TWP PIJF cable with 1cm x 1cm outer insulation damage is shown. 
We pick the three fault detection methods where OMTDR is one of the state-of-the art reflectometry method, S parameter method is more traditional method and the third method uses a widely deployed PLC modems for SNR measurement.
\begin{table}[!b]
       \begin{center}
       \begin{tabular}{|c|c|}
       \hline
       \textbf{Components}& \textbf{Specifications}\\
       \hline
       Computing board & Raspberry PI-3\\
       \hline
           OMTDR sensor & LimeSDR USB (100KHz-3.8GHz)\\
                        
           \hline
           Inductive coupler & Arteche (2-40 MHz)\\
           \hline
           Cable connectors &SMA to BNC connectors\\
           \hline
           Cloud connectivity&4G\\
           \hline
           Cables& i)PIJF 6 quad cable\\
           & ii)PIJF 10 TWP cable\\
           & iii)Symmetric 4 core cables\\
           \hline
           Vector Network& \\
           Analyzer(VNA)& N9923A\\
           \hline
           PLC modem& Devolo dLAN 1200+\\
           \hline
       \end{tabular}
     \caption{ The testbed uses commercially available components and shows their respective specifications}
      \label{tab:testbed_comp}
      \end{center}
    \end{table}
\subsection{OMTDR based Reflectometry test}
The testbed consists of a Raspberry PI-3 (RPi) single board computer  that interfaces with a reflectometry sensor that  monitors the cable health periodically.  The RPi collects the reflectogram data  and communicates the cloud. The reflectometry sensor is  LimeSDR, a  software-defined radio which injects and collects the reflected signal to and from the power line. The transmit (TX) and recieve (RX) ports of the  LimeSDR connect to inductive couplers through coaxial cables. The inductive couplers can easily couple the sensor to the powerline during its normal operation. 
We choose the centre frequency for OMTDR system as 30MHz with a sampling frequency of 5MHz and oversampling rate of 32. We arrived at this combination after several trial rounds and find this optimal for our hardware setup.

In OMTDR based reflectometry test, we inject an OFDM signal onto the CUT and  capture the reflected signal for various load conditions.  We correlate this reflected signal with the transmitted signal and the resultant signal is called a  reflectogram. The correlation peak magnitudes are high at the locations of impedance mismatched cable sections. Every insulation fault introduces an impedance mismatch and part of the input signal is reflected back to the source. We refer to insulation fault as a soft fault. The reflectogram helps us to locate the soft faults and  estimate the severity of those faults. 

\subsubsection{Insulation fault identification and localization using OMTDR}
\begin{figure}
     \centering
     \begin{subfigure}{0.47\textwidth}
         \centering
         \includegraphics[width=0.8\linewidth]{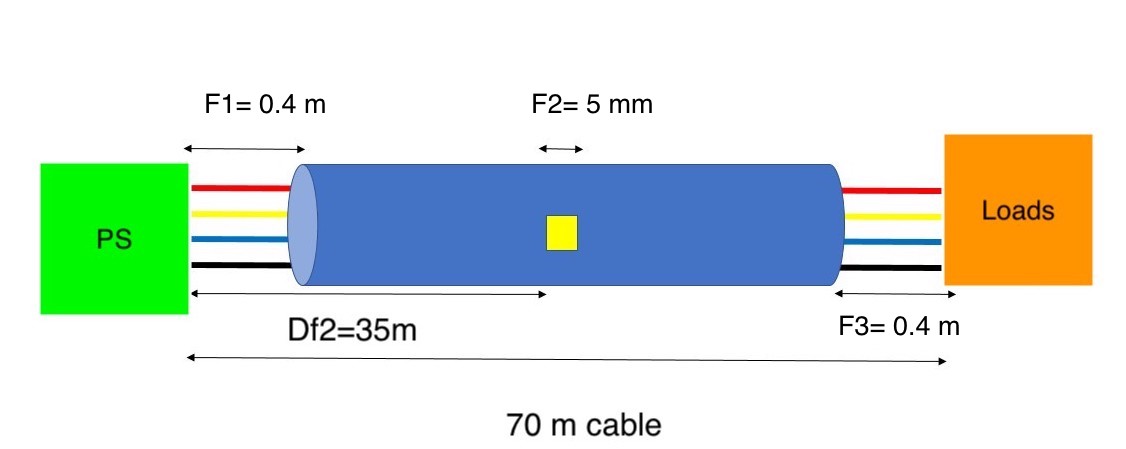}
         \caption{The Cable Under Test (CUT) is connected between Power Supply (PS) and loads. The Fault conditions are F1-fault length of 0.4 m, F2-fault length of 5mm at a distance, Df2 of 35m from PS end and F3- fault length of F3-0.4m at a distance of 69.5m from PS end.}
         \label{fig:longcable_testbed}
     \end{subfigure}
     \hfill
     \begin{subfigure}{0.47\textwidth}
         \centering
         \includegraphics[width=0.8\linewidth]{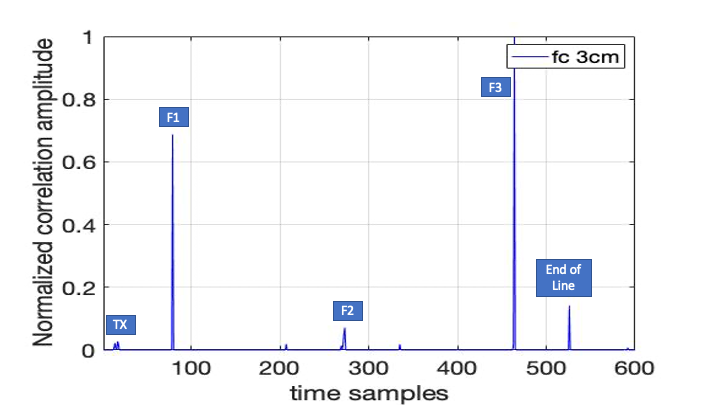}
         \caption{Reflectogram obtained from LimeSDR USB shows CPs corresponding to TX signal, F1, F2, F3 and the end of the line.}
         \label{fig:usb}
     \end{subfigure}
    \caption{Cable fault condition and OMTDR Reflectogram}
        \label{fig:omtdr_results}
        \vspace{-15 pt}
\end{figure}

We analyze the reflectogram to locate the faults. The time samples and the correlation peaks are the two parameters obtained from the reflectogram. The time sample corresponding to the end of the line and transmit signal from baseline measurements helps us to identify the distance of the fault  from the test signal injection. The length of the sample railway cables PIJF 6 quad and PIJF 10TWP cables considered are 24 m and 7.2 m respectively. These cables are connected between 110v 50 Hz PS and a 0.2 KW load. The reflectograms corresponding to these healthy cables are called baseline measurements. We then intentionally introduce an insulation cut of length 1cm and width 1cm at a distance 21 m and 5.79 m in PIJF 6 quad and PIJF 10TWP cables respectively for testing purpose. Reflectogram will have a correlation peak at the time sample corresponding to the fault. Table \ref{tab:PIJF} shows the results, where a fault position error of 0.5 m and 0.17m respectively is seen. 
\begin{table}[!b]
    \begin{center}
    \begin{tabular}{|c|c|c|}
    \hline
    Cable& Actual& Mean measured\\
    type& position& position\\
    \hline
      PIJF   & 21&20.5 \\
       6 quad  && \\
       \hline
       PIJF&5.79&5.62\\
       10 TWP&&\\
       \hline
    \end{tabular}
    \caption{Actual and mean measured position of Insulation fault in PIJF cables using OMTDR method.}
    \label{tab:PIJF}
    \end{center}
\end{table}
  The choice of thresholds for the magnitude of correlation peak at the time sample corresponding to fault, gives a high detection accuracy in addition to the trade off in terms of false positives. 
We also used the testbed for different types of cables with longer lengths. For instance, we would require about 50 to 70m of length to demonstrate faults occuring between two location boxes.  Fig \ref{fig:longcable_testbed} shows a 70m symmetrical four-core conductor connected between the power supply and the loads.  There is a 0.4 m insulation peel off at both ends of the cable. In addition, there is a 5mm insulation fault at a 35m distance from Power Supply (PS). Fig \ref{fig:usb} is the reflectogram collected using LimeSDR USB showing the Correlation Peaks (CP) corresponding to the injected transmitted signal, the three introduced faults, and the end of the line.
 \subsection{PLC-modem based SNR test}
\begin{figure}
    \centering
    \includegraphics[width=0.8\columnwidth]{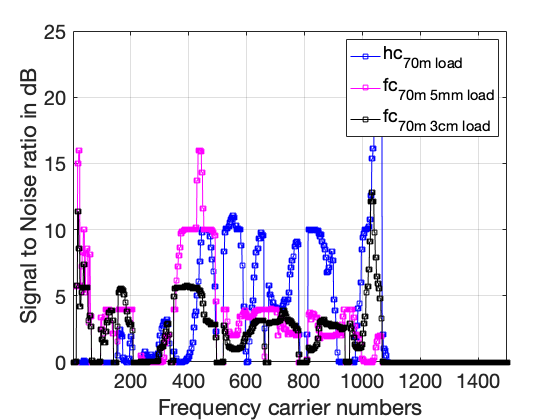}
    \vspace{1pt}
    \caption{Fault identification using SNR measurements using PLC modem. The $hc_{70m load}$ curve represents the average spectral SNR of a healthy cable, while $fc_{70m 5mm load}$ and $fc_{70m 3cm load}$ curves represents the average spectral SNR of a cable with fault length 5mm and 3cm respectively.}
    \label{fig:snr5mm}
\end{figure}

 We use Commercial Off The Shelf (COTS) Power Line Communication (PLC) modems to measure SNR between the two ends of the CUT. These modems have QCA 7500 transceiver chip that responds to the queries from open-plc-utils software \cite{open-plc-utils}. A pair of  PLC modems communicate to each other periodically and calculate the SNR at both ends to estimate the cable conditions. In our lab setup, we connect a CUT between a PS and loads. We inductively couple the two PLC modems at each end of the 70m CUT. We connect each modem to an RPi running open-plc-utils. The RPi measures SNR for varying load conditions. We plot the average spectral SNR obtained by this method as shown in Fig \ref{fig:snr5mm}.  We analyze the SNR plot to know the state of the cable.
\subsection{S parameter test}
\label{sec3c}
\begin{figure}
    \centering
    \includegraphics[width=0.8\columnwidth]{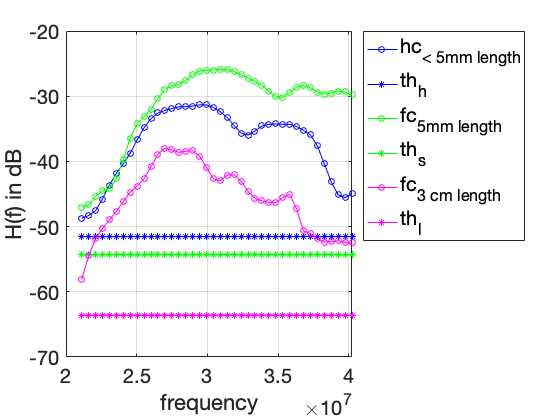}
    \vspace{1pt}
    \caption{Average Channel Frequency Response (CFR) is plotted for various cable conditions. $hc_{<5mm length}$, $fc_{5mm length}$, and $fc_{3cm length}$ curves represents average CFR for the healthy cable, the cable with small fault of size 5mm, the cable with large insulation fault of size 3cm respectively. The straight lines represents the thresholds for the three states of the cable considered.}
    \label{fig:spar_cfr}
    \vspace{-10 pt}
\end{figure}

We measure the S parameters of different cable fault conditions using a VNA. We follow a simple procedure to compute Channel Frequency Response (CFR) from S parameter measurements \cite{NASA2011}. The analysis of this CFR will give us a method to identify the insulation faults. While measuring S Parameters, we vary the load conditions from a full load condition to a no-load condition in a random fashion. We measure the S Parameters for 100 time instances and compute the average CFR from these S parameter measurements. 
We plot the average CFR of the healthy cable, with a $hc_{<5mm length}$ curve shown in Fig \ref{fig:spar_cfr}. While we plot the $fc_{5mm length}$ and $fc_{3cm length}$ curves for the average CFR of the cables with 5mm and 3cm insulation cuts. The mean of each curve with phase information gives a  threshold level for healthy, small fault, and large fault cases. Later these threshold values are used for online measurements to estimate the state of an unknown cable condition. 

\section{Probability of cable fault detection}
We discuss the probability of identifying the state of the cable by the S parameter-based measurement method. This can be easily generalized for any other measurement method. We explain the procedure to obtain the conditional probability of the actual state of the CUT given the measurement-based estimated state of the CUT. Initially, we classify a CUT based on the length of the insulation cut. If a cable has an insulation fault length greater than 3cm, we classify it as a large fault, $F_l$. Similarly, if the insulation fault length is between 5mm and 3cm, we classify it as a small fault, $F_s$ and the cable with insulation fault length less than 5mm as healthy cable. We perform several measurements on known cable states and use this information to set thresholds to identify each of these states. We use Bayesian inversion framework to use the knowledge of estimated state of the cable to infer the actual state of the cable.
When we do an S Parameter measurement on a cable, three events such as small fault $F_s$, large fault, $F_s$, and no-fault $H$ can occur. Let $Y_1$ represent the estimated state of the CUT by a measurement method and $S$ represent the actual state of the  CUT. Let $Y_1$ and $S$ take any one of the mutually exclusive states, healthy ($H$), small insulation cut ($F_s$), and large insulation cut ($F_l)$. The probability of the event $Y_1$ taking a particular state, for example, $P(Y_1=F_l)$ is the sum of probabilities of the three events as in equation 
\ref{y1 eq}.
\begin{equation}
\label{y1 eq}
    \mathtt{P(Y_1=F_l)}=\mathtt{\sum_{s_1 \in (F_s,F_l,H)} P((Y_1=F_l),(S=s_1))}
\end{equation}
The joint probability $P((Y_1=F_l),(S=s_1))$ can then be represented as
\begin{equation}
\mathtt{P((Y_1=F_l),(S=s_1))}=\mathtt{P((Y_1=F_l)/(S=s_1))P(S=s_1)}
\end{equation}
where $P(S=s_1)$ is the a priori probability of the actual state of the CUT.
We then obtain the conditional probability
\begin{align} \label{eq:cond_prob_final}
   \mathtt{P((S=s_1)/(Y_1=F_l))}=\mathtt{\frac{P((Y_1=F_l),(S=s_1))}{P(Y_1=F_l)}}
\end{align}
where $P((S=s_1)/(Y_1=s_1))$ is the conditional probability of the actual state of the CUT given the estimated state of the CUT. We repeat the same computation for all values of $s_1 \in (F_l,F_s,H).$ The value obtained in~\eqref{eq:cond_prob_final} provides a trustworthiness metric of the considered method. We explain the procedure to obtain these conditional probabilities empirically using Algorithm \ref{cond_prob_sbyy1}. We measure S parameters with a VNA for the healthy cable, the cable with 5mm, and  the cable with 3cm insulation cuts.  We measure average CFR for $N_1$ time instants for all three cable conditions. We estimate the values for threshold settings by the procedure mentioned in section \ref{sec3c} and use those values in Algorithm \ref{cond_prob_sbyy1}. 
The $\psi_i$ parameters used in Algorithm \ref{cond_prob_sbyy1} corresponds to average SNR, average CFR, and average CP for SNR test, S parameter test and OMTDR test respectively. We use appropriate parameters and the corresponding thresholds for each test. $Y_1$, $Y_2$, and $Y_3$ represents  the Estimated state of the CUT by S parameter, SNR, and OMTDR measurements respectively. 
\subsection{Health Index}
To quantify the health and remaining operating life of a cable, we utilize a metric called the health index. Health index quantifies the quality of the cable with a low value indicating a poor health cable condition and a high value indicating a healthy condition.
\begin{algorithm}[!b]
\small
 \SetAlFnt{\small}
\SetAlgoLined
{
Input $s_1 \in {F_l,F_s,H}$\;
Input $P(S=H), P(S=F_s), P(S=F_l)$\;
$i=1$\;

}
 \While{$i \leq N_1$}{
 {
 $LF_i=0$\;
 $SF_i=0$\;
 $H_i=0$\;
 }
 \uIf{$\psi_i \geq th_l$}{
  $LF_i=1$\;
   }
 \uElseIf{$th_l \leq \psi_i\leq th_s$}{
  $SF_i=1$\;
   }
  \Else{
  $H_i=1$\;
  }
  $i=i+1$\;
 }
$P\Big(\nicefrac{(Y_1=F_l)}{(S=s_1)}\Big)=\frac{\sum_{i=1}^{N_1}LF_i}{N_1}$\;
 $P\Big(\nicefrac{(Y_1=F_s)}{(S=s_1)}\Big)= \frac{\sum_{i=1}^{N_1} SF_i}{N_1}$\;
$P\Big(\nicefrac{(Y_1=H)}{(S=s_1)}\Big)= \frac{\sum_{i=1}^{N_1} H_i}{N_1}$\;
$P(Y_1=F_l,S=s_1)=P(\nicefrac{Y_1=F_l}{S=s_1})\cdot P(S=s_1)$\;
 $P(Y_1=F_l)=\underset{s_1 \in {F_l,F_s,H}}{\sum}P(Y_1=F_l,S=s_1)$\;
 $P(\nicefrac{(S=F_l)}{(Y_1=F_l)})= \frac{P(Y_1=F_l,S=F_l)}{P(Y_1=F_l)}$\;
 \caption{Calculation of $P\Big(\frac{S}{Y_1}\Big)$}
 \label{cond_prob_sbyy1}
\end{algorithm}
\setlength{\textfloatsep}{0.1cm}
\setlength{\floatsep}{0.1cm}
\begin{equation}
\small
\begin{aligned}
\label{HI eq}
CFD&=\textstyle{W_1\cdot\frac{\sum_{i=1}^{N_1}Y_{1,i}}{N_1}+W_2\cdot\frac{\sum_{i=1}^{N_2} Y_{2,i}}{N_2}}
       +W_3\cdot\frac{\sum_{i=1}^{N_3} Y_{3,i}}{N_3}\\
 NCFD&=\frac{CFD}{W_1+W_2+W_3}\\
 HI&= (1-NCFD)\cdot 100
 \end{aligned}
\end{equation}
 $W_1, W_2$, and $W_3$ gives an overall trustworthiness metric for each fault detection method. This takes into account of false positives and true positives in identifying the states of the cable using Algorithm \ref{cond_prob_sbyy1}. Weights, $W_i$ for $i \in {1,2,3}$ is computed using Eq \ref{W eq}. $Y_{1,i}$ is the $i^{th}$ measurement result in S-parameter test and it takes value 1 when $|avgCFR|> th_l$ otherwise value 0 is assigned. Similarly $Y_{2,i}$ and $Y_{3,i}$ are results from SNR and OMTDR tests. We then calculate 
Normalized Coefficient of Fault Detection, NCFD using all three methods and eventually provide the composite health index, HI for the cable. $N_1$, $N_2$, and $N_3$ are the total number of time instants at which each experiment is repeated using VNA, PLC modem, and OMTDR sensor respectively. We use  60\% of lab measurements from each test for weight computation and 40\% of measurements for fault detection. 
\begin{equation}
\begin{aligned}
 \label{W eq}
     W_i=\alpha \cdot  P \Big(\nicefrac{(S=F_l)}{(Y_i=F_l)}\Big) + \beta \cdot  P \Big(\nicefrac{(S=F_s)}{(Y_i=F_s)}\Big)\\+\gamma \cdot  P \Big(\nicefrac{(S=H)}{(Y_i=H)}\Big)\\
     \alpha+\beta+\gamma=1
\end{aligned}
\end{equation}
\begin{figure}
    \centering
    \includegraphics[width=0.9\columnwidth]{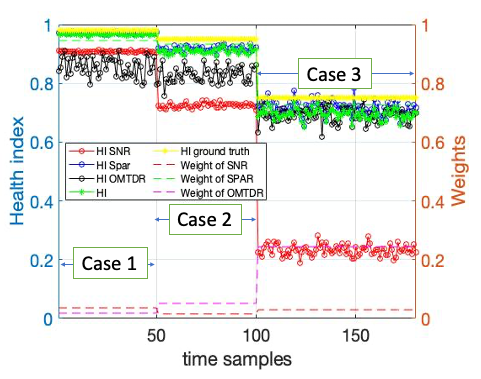}
    \vspace{1 pt}
    \caption{Health index using individual and combined fault detection methods.SNR, S Parameter, OMTDR measurement methods evaluate fault detection. The thresholds for each method is chosen to \textbf{prioritize fault detection} and weights for each method are displayed at the right yaxis.}
    \label{fig:HI_high_det}
    \vspace{-5 pt}
\end{figure}
\begin{figure}
    \centering
    \includegraphics[width=0.9\columnwidth]{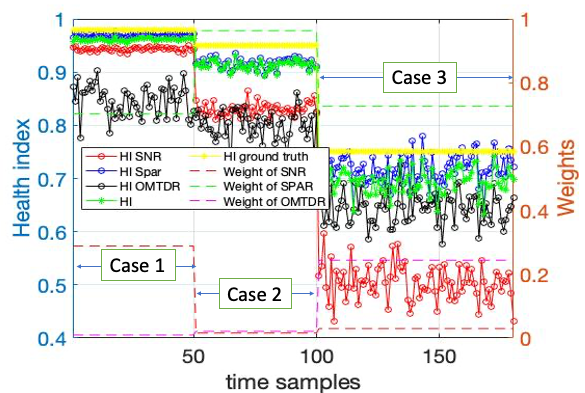}
    \vspace{1 pt}
    \caption{Health index using individual and combined fault detection methods. The thresholds are chosen to \textbf{prioritize low false positives}}
    \label{fig:HI_low_fp}
    \vspace{-5 pt}
\end{figure}
Let us consider three cases to study the significance of health index calculation that employs all three fault detection methods. \textbf{Case1:} The cable with 90\% healthy, 8\% small faults and 2\% large faults chances. This case is emulated with measurement results chosen with same percentages from three states of the cable. \textbf{Case2:} The cable with 17\% healthy, 58\% small faults and 25\% large faults chances. \textbf{Case3:} The cable with 70\% healthy, 25\% small faults and 5\% large faults chances.  Fig \ref{fig:HI_high_det} shows the HI estimated by individual and composite method while prioritizing the fault detection probability. The thresholds are chosen accordingly. Fig \ref{fig:HI_low_fp} shows the estimated HI while prioritizing low false positives. The time samples 0-50, 50-100, and 100-180 corresponds to the case1, case2, and case3 respectively as shown in Fig \ref{fig:HI_high_det} and \ref{fig:HI_low_fp}. The end user has the flexibility in  the level of prioritizing fault detection capability and low false positives by tuning the values of $\alpha$, $\beta$, and $\gamma$ in equation \ref{W eq}.
\begin{equation}
\small
\begin{aligned}
\label{ind_HI}
HI_{Spar}=1-\frac{\sum_{i=1}^{N_1} Y_{1,i}}{N_1}\\
HI_{SNR}=1-\frac{\sum_{i=1}^{N_2} Y_{2,i}}{N_2}\\
HI_{OMTDR}=1-\frac{\sum_{i=1}^{N_3} Y_{3,i}}{N_3}\\
\end{aligned}
\end{equation} 
We calculate the individual HI for each method using equation set \ref{ind_HI}.  
Later we compare the individual and composite HI with the ground truth HI of the cable. The composite HI is closer to the ground truth as the weights to each of the combined methods is chosen from previous measurement inferences.
\section{Conclusion}
We provide an IoT solution for cable health monitoring for track-side point machine, signal post, and train on track detection circuits. We suggest three methodologies to identify insulation fault in signal and power cables associated with these electrical circuits. These methods support online and non-invasive monitoring, and the measurements are performed on a live wire when the system is under normal operation. We analyze the resulting reflectogram, channel frequency response, and SNRs for possible faults in the wire. The Composite Health index calculation using these methods quantifies cable health  in a reliable manner.   
\section*{Acknowledgment}
The authors would like to thank Southeastern  Indian Railways for providing the sample cables for research and their interesting discussions.
\ifCLASSOPTIONcaptionsoff
  \newpage
\fi
\bibliographystyle{IEEEtran}
\bibliography{ref}
\end{document}